\documentclass[11pt,draftcls,onecolumn]{IEEEtran}

\usepackage{amsfonts}
\usepackage{amssymb}
\usepackage{mathrsfs}
\usepackage{amsmath}
\usepackage{cite}
\usepackage{mathrsfs}
\usepackage[ruled,vlined]{algorithm2e}
\usepackage{tikz}
\usetikzlibrary{arrows,automata}
\usepackage[latin1]{inputenc}
\usepackage{verbatim}
\makeatletter

\newcommand{\Rmnum}[1]{\expandafter\@slowromancap\romannumeral #1@}
\makeatother

\newtheorem{thm}{Theorem}
\newtheorem{lemma}[thm]{Lemma}
\newtheorem{eg}{Example}
\newtheorem{prop}{Proposition}
\newtheorem{cor}[thm]{Corollary}
\newtheorem{defn}{Definition}
\newtheorem{rem}[thm]{Remark}

\newtheorem{rem-eg}[thm]{Remark and Example}

\newcommand{\ti}{\tilde}

\newcommand{\Dt}{\Delta_t}

\newcommand{\w}{{\omega}}
\newcommand{\bX}{{\bf X}}

\newcommand{\bZ}{{\bf Z}}
\newcommand{\p}{{\rho}}
\newcommand{\dt}{{\delta_t}}

\newcommand{\f}{\tilde{f}}
\newcommand{\g}{\tilde{g}}
\newcommand{\F}{\tilde{F}}
\newcommand{\bzero}{{\bf 0}}
\newcommand{\bg}{{\bf g}}
\newcommand{\bk}{{\bf k}}

\newcommand{\mL}{\mathcal{L}}
\newcommand{\mF}{\mathcal{F}}

\newcommand{\mO}{\mathcal{O}}
\newcommand{\mE}{\mathcal{E}}
\newcommand{\mC}{\mathbf{C}}
\newcommand{\Rank}{{\mathrm{Rank}}}
\newcommand{\vk}{\vec{k}}
\newcommand{\rt}{{\rm row}_t}

\begin{document}

\title{Variable-Rate Linear Network Error Correction MDS Codes
\thanks{
This research is supported by the National Key Basic Research Problem of China (973 Program Grant No. 2013CB834204), the National Natural Science Foundation of China (Nos. 61301137, 61171082) and the Fundamental Research Funds for Central Universities of China (No. 65121007).
The material in this paper was presented in part at the IEEE International Symposium on Network Coding, Beijing, China, July 2011.}}
\author{Xuan~Guang,
        Fang-Wei~Fu,
        and~Zhen Zhang,~\IEEEmembership{Fellow,~IEEE}
\thanks{X. Guang is with the School of Mathematical Science and LPMC, Nankai University, Tianjin 300071, P. R. China. Email:
xguang@nankai.edu.cn.}
\thanks{F.-W. Fu is with the Chern Institute of
Mathematics and LPMC, Nankai University, Tianjin 300071, P. R. China. Email:
fwfu@nankai.edu.cn.}
\thanks{Z. Zhang is with the Communication Sciences Institute, Ming Hsieh Department of
Electrical Engineering, University of Southern California, Los Angeles,
CA 90089-2565 USA. Email: zhzhang@usc.edu.}}

\markboth{Variable-Rate Linear Network Error Correction MDS Codes}%
{Guang \MakeLowercase{\textit{et al.}}: Universal Linear Network Error Correction MDS Codes}
%


\maketitle

\begin{abstract}
In network communication, the source often transmits messages at several different information rates within a session. How to deal with information transmission and network error correction simultaneously under different rates is introduced in this paper as a variable-rate network error correction problem. Apparently, linear network error correction MDS codes are expected to be used for these different rates. For this purpose, designing a linear network error correction MDS code based on the existing results for each information rate is an efficient solution. In order to solve the problem more efficiently, we present the concept of variable-rate linear network error correction MDS codes, that is, these linear network error correction MDS codes of different rates have the same local encoding kernel at each internal node. Further, we propose an approach to construct such a family of variable-rate network MDS codes and give an algorithm for efficient implementation. This approach saves the storage space for each internal node, and resources and time for the transmission on networks. Moreover, the performance of our proposed algorithm is analyzed, including the field size, the time complexity, the encoding complexity at the source node, and the decoding methods. Finally, a random method is introduced for constructing variable-rate network MDS codes and we obtain a lower bound on the success probability of this random method, which shows that this probability will approach to one as the base field size goes to infinity.
\end{abstract}

%
\IEEEpeerreviewmaketitle

\section{Introduction}
\IEEEPARstart{N}{etwork} coding allows internal nodes in a communication network to process the information received. This idea was first appeared in Yeung and Zhang \cite{Zhang-Yeung-1999} and then developed by Ahlswede \textit{et al.} \cite{Ahlswede-Cai-Li-Yeung-2000}. In \cite{Ahlswede-Cai-Li-Yeung-2000}, the authors showed that if coding is applied at the nodes in a network, rather than routing alone, the source node can multicast messages to all sink nodes at the theoretically maximum rate---the smallest minimum cut capacity between the source and any sink node, as the alphabet size approaches infinity. Li \textit{et al.} \cite{Li-Yeung-Cai-2003} further indicated that linear network coding with finite alphabet size is
sufficient for multicast. Koetter and M\'{e}dard \cite{Koetter-Medard-algebraic} developed an algebraic characterization of network coding. Although network coding has higher information rate than classical routing, Jaggi \textit{et al.} \cite{co-construction} still proposed a deterministic polynomial-time algorithm for constructing a linear network code. For a detail and comprehensive discussion of network coding, refer to \cite{Zhang-book, Yeung-book, Fragouli-book, Ho-book}.

Network coding has been studied extensively for several years under the assumption that the channels of networks are error-free. Unfortunately, all kinds of errors may occur in practical network communication such as random errors, erasure errors (packet losses), errors in headers and so on. In order to deal with such problems, network error correction (NEC) based on network coding was studied recently.  Cai and Yeung proposed the original idea of network error correction coding in their conference paper \cite{Yeung-Cai-coorrect} and developed it in their journal papers \cite{Yeung-Cai-correct-1}\cite{Yeung-Cai-correct-2}. They introduced the concept of network error correction codes as a generalization of classical error-correcting codes, and extended some important bounds in classical coding theory to network error correction coding, such as the Singleton bound, the Hamming bound, and the Gilbert-Varshamov bound. Although the Singleton bound has been given by Yeung and Cai \cite{Yeung-Cai-correct-1}, Zhang\cite{zhang-correction} and Yang \textit{et al.} \cite{Yang-refined-Singleton}\cite{Yang-thesis} presented the refined Singleton bound independently by the different approaches. Further, the linear NEC codes satisfying this bound with equality are called linear network error correction maximum distance separable (MDS) codes, or network MDS codes for short.

Koetter and Kschischang \cite{Koetter-correction} (see also \cite{Silva-K-K-rank-metric-codes}) formulated a different framework for network error correction coding when a noncoherent network model was under consideration where neither source node nor sink node was assumed to have knowledge of the channel transfer characteristic. Motivated by the property that linear network coding is vector-space preserving, in their approach the source message is represented by a
subspace of a fixed vector space and a basis of the subspace is injected into the network. So this type of network error correction codes is called subspace codes. A metric was proposed to account for the discrepancy between the transmitted and received subspaces and a coding theory based on this metric was developed. For an overview of the development and some contributions in network error correction coding, refer to the survey paper \cite{Zhang-survey-paper-NEC}.

In network communication, the source often transmits the messages at several different information rates within a session. When both information transmission and network error correction are considered simultaneously, it is expected that linear network error correction MDS codes can be applied for these information rates. For the problem as described above, the most efficient solution based on the existing results is that, for each information rate, design a network MDS code by constructive algorithms proposed by Yang \textit{et al.} \cite{Yang-refined-Singleton},  Guang \textit{et al.} \cite{Guang-MDS}, or others. For this scheme, each node in a network has to store all local encoding kernels corresponding to different network MDS codes. Hence, it takes a large amount of storage space for each node in network. This also increases the complexity of the system considerably. Furthermore, in transmission, the source node has to tell each non-source node which information rate is used to transmit the messages, and then each non-source node searches and uses the corresponding local encoding kernel for coding. Searching and changing the local encoding kernels at each non-source node consume resources and time in the network.

In order to avoid these shortcomings of the above solution, we wish to construct a family of linear network error correction MDS codes with the following property: these network MDS codes with different rates have the same local encoding kernel at each non-source node. In other words, for these different information rates, each non-source node can use the same local encoding kernel for coding. This will save the storage space at each node and all internal nodes will not need to know which rate the source node uses to transmit messages. We are partly motivated by the same problem in network coding \cite{Fong-Yeung-variable-rate}, where, Fong and Yeung studied the variable-rate linear network coding with and without link failure in the case that no errors occur in the channels of networks. \cite{Si-V-R-NetCod11} and \cite{Sun-V-R-NetCod12} further studied different classes of variable-rate linear network codes.

This paper is divided into 6 sections. In the next section, we first review linear network coding and linear network error correction coding, and then give some necessary notation and definitions. Section \Rmnum{3} is devoted to constructing variable-rate linear network error correction MDS codes and designing an algorithm for efficient implementation. In this section, we give a method to construct low-dimensional linear network MDS codes from a high-dimensional one such that they have the same local encoding kernel at each non-source node. In other words, we give a constructive proof to show the existence of the variable-rate network MDS codes defined in Section \Rmnum{3}. Actually, the existence may be proved more easily by a random method as used in \cite{zhang-random}\cite{Guang-aver-prob-IEICE}. But the constructive approach is much more important because of its widely potential applications. Furthermore, we design an algorithm for efficient implementation. Section \Rmnum{4} is devoted to the performance analysis of our proposed algorithm in Section \Rmnum{3}, including the field size, the time complexity of the algorithm, the encoding complexity at the source node, and the decoding methods. In particular, we also discuss the feasibility of two algorithms proposed by Yang \textit{et al.} \cite{Yang-refined-Singleton} for this variable-rate network error correction problem. Since both algorithms design the codebook at the source node and the local encoding kernels separately, it seems likely that they might solve this variable-rate problem. However, by a detailed analysis, they are either non-feasible or inefficient for solving the problem. Particularly, even assuming that some certain conditions are satisfied such that Algorithm 1 in \cite{Yang-refined-Singleton} can solve our problem, our proposed algorithm still has many advantages in different aspects. In Section \Rmnum{5}, a random approach for implementing variable-rate network error correction MDS codes is proposed and then we obtain a lower bound on the success probability of using the random approach to construct variable-rate network MDS codes. This success probability can characterize the performance of this random method, and the obtained lower bound implies that, if the field size is sufficiently large, the random method can construct variable-rate network MDS codes with high probability close to one. The last section summarizes the works done in this paper and proposes some topics for further research.

\section{Preliminaries}
In the present paper, we follow \cite{zhang-correction}\cite{Guang-MDS} with their notation and terminology.
A communication network is represented as a finite acyclic directed
graph $G=(V,E)$, where $V$ and $E$ are the sets of
nodes and channels of the network, respectively. The node set $V$ consists of three
disjoint subsets $S$, $T$, and $J$, where $S$ is the set of source
nodes, $T$ is the set of sink nodes, and $J=V-S-T$ is
the set of internal nodes. A direct edge $e=(i,j)\in E$
stands for a channel leading from node $i$ to node $j$. Node $i$
is called the tail of $e$ and node $j$ is called the
head of $e$, denoted by $tail(e)$ and $head(e)$, respectively. Correspondingly, the channel $e$ is
called an outgoing channel of $i$ and an incoming channel of $j$. For a node $i$, define $Out(i)$ as the set of outgoing channels of $i$ and $In(i)$ as the set of incoming channels of $i$. Formally, we have
\begin{align*}
Out(i)=\{e\in E:\ tail(e)=i\},\ \ \ In(i)=\{e \in E:\ head(e)=i\}.
\end{align*}
For each channel $e\in E$, there exists a
positive number $R_e$, say the capacity of $e$. We allow the multiple channels between two nodes and
thus assume reasonably that the capacity of any channel is 1 per unit time, that is, one field element can be transmitted over a channel in one unit time. A cut between node $i$ and node $j$ is a set of channels whose removal disconnects $i$ from $j$. For unit capacity channels, the capacity of a cut can be regarded as the number of channels in the cut, and the minimum of all capacities of cuts between $i$ and $j$ is called the minimum cut capacity between the two nodes. A cut between node $i$ and node $j$ is called a minimum cut if its capacity achieves the minimum cut capacity between them. Note that there may exist several minimum cuts between $i$ and $j$, but the minimum cut capacity between them is determined. Following the direction of the channels, there is an upstream-to-downstream order (ancestral topological order) on the channels in $E$ which is consistent with the partial order of all channels. The coordinates of all vectors and rows/columns of all matrices in this paper are indexed according to this upstream-to-downstream order. In particular, if $L$ is such a matrix whose column vectors are indexed by a collection $B\subseteq E$ of channels according to an upstream-to-downstream order, then we use some symbol with subscript $e$, $e\in B$, such as $l_e$, to denote the column vector indexed by the channel $e$, and the matrix $L$ is written as column-vector form $L=\Big[l_e:\ e\in B\Big]$. If $L$ is a matrix whose row vectors are indexed by this collection $B$ of channels, then we use some symbol with $e$ inside a pair of brackets, such as $l(e)$, to denote the row vector corresponding to $e$, and the matrix $L$ is written as row-vector form $L=\Big[ l(e):\ e\in B \Big]$.

\subsection{Linear Network Coding}

In this paper, we consider single source networks, i.e., $|S|=1$, and the unique source node is denoted by $s$, which generates messages and transmits them to all sink nodes over the network by a linear network code. The source node $s$ has no incoming channels and any sink node has no outgoing channels. But we introduce the concept of imaginary incoming channels of the source node $s$ and assume that these imaginary incoming channels provide the source messages to $s$. Let the information rate be $\w$ symbols per unit time. Then $s$ has $\w$ imaginary incoming channels denoted by $d_1',d_2',\cdots,d_\w'$ and let $In(s)=\{d_1',d_2',\cdots,d_\w'\}$. The source messages are $\w$ symbols $\bX=[X_1\ X_2\ \cdots\ X_\w]$ arranged in a row vector where each $X_i$ is an element of the base field $\mF$. Subsequently, they are assumed to be transmitted to $s$ through the $\w$ imaginary incoming channels in $In(s)$. Without loss of generality, assume that the message transmitted over the $i$th imaginary channel is the $i$th source message. Further, at each node $i\in V-T$, there is an $|In(i)|\times|Out(i)|$ matrix $K_i=[k_{d,e}]_{d\in In(i),e\in Out(i)}$, called the local encoding kernel at $i$, where $k_{d,e}\in \mF$ is called the local encoding coefficient for the adjacent pair $(d,e)$ of channels. We use $U_e$ to denote the message transmitted over the channel $e$. Hence, at the source node $s$, we have $U_{d_i'}=X_i$, $1\leq i \leq \w$. In general, the message $U_e$ transmitted over the channel $e\in E$ is calculated recursively by the formulae: $$U_e=\sum_{d\in In(tail(e))}k_{d,e}U_d.$$
Furthermore, it is not difficult to see that $U_e$ is actually a linear combination of the $\w$ source symbols $X_i$, $1\leq i\leq \w$, that is, there is an $\w$-dimensional column vector $f_e$ over the base field $\mF$ such that $U_e=\bX \cdot f_e$ (see also \cite{Zhang-book} \cite{Yeung-book}). This column vector $f_e$ is called the global encoding kernel of a channel $e$, and can be determined by the local encoding kernels as follows:
$$f_e=\sum_{d\in In(tail(e))}k_{d,e}f_d,$$
with boundary condition that the vectors $f_{d_i'}$, $1\leq i \leq \w$, form the standard basis of the vector space $\mF^\w$.

\subsection{Linear Network Error Correction Coding}

In the case that an error occurs on a channel $e$, the output of the channel is $\tilde{U}_e=U_e+Z_e$, where $U_e$ is the message that should be transmitted over the channel $e$ and $Z_e\in \mF$ is the error occurred in $e$. We also treat the error $Z_e$ as a message called \textit{error message}. Further, let the error vector be an $|E|$-dimensional row vector $\bZ=[Z_e:\ e\in E]$ over the field $\mF$ with each component $Z_e$ representing the error occurred on the corresponding channel $e$. Firstly, we introduce the extended network as follows. In the network $G=(V,E)$, for each channel $e\in E$, an imaginary channel $e'$ is introduced, which is connected to the tail of $e$ in order to provide the error message $Z_e$. This new network $\tilde{G}=(\tilde{V},\tilde{E})$ with imaginary channels is called the extended network of $G$, where $\tilde{V}=V$, $\tilde{E}=E\cup E'\cup \{d_1',d_2',\cdots, d_\w'\}$ with $E'=\{e': e\in E\}$. Obviously, $|E'|=|E|$. Then a linear network code for the original network $G$ can be extended to a linear network code for the extended network $\tilde{G}$ by setting $k_{e',e}=1$ and $k_{e',d}=0$ for all $d\in E\backslash\{e\}$. Note that, for each internal node $i$ in the extended network $\tilde{G}$, $In(i)$ only includes the real incoming channels of $i$, that is, the imaginary channels $e'$ corresponding to $e\in Out(i)$ are not in $In(i)$. But for the source node $s$, we still define $In(s)=\{d_1',d_2',\cdots,d_\w'\}$. In order to distinguish two different types of imaginary channels, we say $d_i'$, $1\leq i\leq \w$, the \textit{imaginary message channels} and $e'$ for $e\in E$ the \textit{imaginary error channels}. Similarly, we can also define global encoding kernels $\f_e$ for all $e\in \tilde{E}$, which is an $(\w+|E|)$-dimensional column vector and the entries can be indexed by the channels in $In(s)\cup E$. For imaginary message channels $d_i'$, $1\leq i \leq \w$, and imaginary error channels $e'\in E'$, let $\f_{d_i'}=1_{d_i'}$ and $\f_{e'}=1_e$, where $1_d$ is an $(\w+|E|)$-dimensional column vector which is the indicator function of $d\in In(s)\cup E$. Thus, the vectors $\f_e$ for both $\w$ imaginary message channels and $|E|$ imaginary error channels form the standard basis of vector space $\mF^{\w+|E|}$. For other global encoding kernels $\f_e$, $e\in E$, we have the following recursive formulae:
$$\f_e=\sum_{d\in In(tail(e))}k_{d,e}\f_d+1_e.$$
We call $\f_e$ the extended global encoding kernel of the channel $e$ for the original network. At each sink node $t\in T$, the received message vector $\ti{U}_t\triangleq [\ti{U}_e:\ e\in In(t)]$ and the decoding matrix $\F_t\triangleq \begin{bmatrix}\f_e:\ e\in In(t)\end{bmatrix}$ are available, and we have the following decoding equation:
\begin{align*}
\ti{U}_t=(\bX\ \bZ)\F_t,
\end{align*}
which can be used for decoding and error correction (refer to \cite{zhang-correction}\cite{Guang-MDS}).

Similar to linear network codes \cite{Zhang-book}\cite{Yeung-book}, we can also define a linear network error correction code by either a local description or a global description.
\begin{defn}\
\begin{description}
  \item [{\bf Local Description of A Linear Network Error Correction Code.}]\

   An $\w$-dimensional $\mF$-valued linear network error correction code consists of all local encoding kernels at all internal nodes (including the source node $s$), i.e., $$ K_i=[k_{d,e}]_{d\in In(i), e\in Out(i)},$$
   that is an $|In(i)|\times |Out(i)|$ matrix for the node $i$, where $k_{d,e}\in \mF$ is the local encoding coefficient for the adjacent pair $(d,e)$ of channels with $d\in In(i)$, $e\in Out(i)$.
  \item[{\bf Global Description of A Linear Network Error Correction Code.}]\

  An $\w$-dimensional $\mF$-valued linear network error correction code consists of all extended global encoding kernels for all channels including imaginary message channels and imaginary error channels, which satisfy:
      \begin{enumerate}
        \item $\f_{d_i'}=1_{d_i'},\ 1 \leq i \leq \w$, and $\f_{e'}=1_e$, $e'\in E'$, where $1_d$ is an $(\w+|E|)$-dimensional column vector which is the indicator function of $d\in In(s) \cup E$;
        \item for other channels $e\in E$,
        \begin{align}\label{equ_ext_f}
        \f_e=\sum_{d\in In(tail(e))}k_{d,e}\f_d+1_e,
        \end{align}
        where $k_{d,e}\in \mF$ is the local encoding coefficient for the adjacent channel pair $(d,e)$ with $d\in In(tail(e))$, and again $1_e$ is an $(\w+|E|)$-dimensional column vector which is the indicator function of the channel $e\in E$.
      \end{enumerate}
\end{description}
\end{defn}

Further, we give the following notation and definitions.

\begin{defn}
For each channel $e\in E$, the extended global encoding kernel $\f_e$ is written as follows:
\begin{equation*}
\f_e=\begin{bmatrix}
f_e(d_1')\\
\vdots\\
f_e(d_\w')\\
f_e(e_1)\\
\vdots\\
f_e(e_{\mE})\\
\end{bmatrix}
=\begin{bmatrix}
f_e\\
g_e\\
\end{bmatrix}
\end{equation*}
where
$f_e=\begin{bmatrix}
f_e(d_1')\\
\vdots\\
f_e(d_\w')
\end{bmatrix}
$ is an $\w$-dimensional column vector, and
$g_e=\begin{bmatrix}
f_e(e_1)\\
\vdots\\
f_e(e_{\mE})\\
\end{bmatrix}$
is an $\mE$-dimensional column vector with $|E|=\mE$.
\end{defn}

Recall that $\F_t=\begin{bmatrix}\f_e:\ e\in In(t)\end{bmatrix}$ is the decoding matrix at the sink node $t\in T$. Denote by $\rt(d)$ the row vector of the decoding matrix $\F_t$ indexed by the channel $d\in In(s)\cup E$. These row vectors are of dimension $|In(t)|$. Hence,
\begin{equation*}
\tilde{F}_t=\begin{bmatrix}
\rt(d_1')\\
\vdots\\
\rt(d_\w')\\
\rt(e_1)\\
\vdots\\
\rt(e_{\mE})
\end{bmatrix}
=
\begin{bmatrix}
F_t\\
G_t
\end{bmatrix}
\end{equation*}
where $F_t=\begin{bmatrix}
\rt(d_1')\\
\vdots\\
\rt(d_\w')\end{bmatrix}$ and $G_t=\begin{bmatrix}\rt(e_1)\\ \vdots \\ \rt(e_{\mE}) \end{bmatrix}$ are two matrices of sizes $\w\times |In(t)|$ and $|E|\times |In(t)|$, respectively.

We use $\p$ to denote an error pattern which can be regarded as a set of channels. We say that an error message vector $\bZ$ matches an error pattern $\p$, if $Z_e=0$ for all $e\in E\backslash \p$. In the following, we always use $\bzero$ to denote an all zero row vector, whose dimension will always be clear from the context.

\begin{defn}[{\cite[Defintion 3]{zhang-correction}}]
Define
$$\Delta(t,\p)=\{(\bzero\ \bZ)\ti{F}_t=\bZ\cdot G_t:\ \bZ\in \mF^{|E|} \mbox{ matching the error pattern }\p \},$$
and
$$\Phi(t)=\{(\bX\ \bzero)\ti{F}_t=\bX\cdot F_t:\ \bX\in \mF^\w \}.$$
We call $\Delta(t,\p)$ and $\Phi(t)$ the error space of the error pattern $\p$ and the message space with respect to the sink node $t$, respectively.
\end{defn}

Let $L$ be a collection of vectors in some linear space. For convenience, we use $\langle L \rangle$ to represent the subspace spanned by vectors in $L$. Thus, we further have
\begin{align*}
\Delta(t,\p)=\langle \{ \rt(d):\ d\in \p \} \rangle \mbox{ and }  \Phi(t)=\langle \{ \rt(d):\ d\in In(s) \} \rangle.
\end{align*}

Moreover, we give some concepts which will be used in this paper.

\begin{defn}[{\cite[Definition 4]{zhang-correction}}]
We say that an error pattern $\p_1$ is dominated by another error pattern $\p_2$ with respect to a sink node $t$, if $\Delta(t,\p_1)\subseteq \Delta(t,\p_2)$ for any linear network code. This relation is denoted by $\p_1\prec_t\p_2$.
\end{defn}

\begin{defn}[{\cite[Definition 5]{zhang-correction}}]
The rank of an error pattern $\p$ with respect to a sink node $t$ is defined by
$$rank_t(\p)=\min\{|\p'|:\ \p\prec_t\p'\},$$
where $|\p'|$ denotes the cardinality of the error pattern $\p'$.
\end{defn}

The above definition on the rank of an error pattern is abstract, and so in order to understand this concept more intuitively, we give the following proposition.
\begin{prop}[{\cite[Proposition 1]{Guang-MDS}}]\label{prop_error_pattern}
For an error pattern $\p$, introduce a source node $s_{\p}$. Let $\p=\{ e_1,e_2,\cdots,\ e_l \}$ where $e_j\in In(i_j)$ for $1\leq j \leq l$ and define new edges $e_j'=(s_{\p},i_j)$. Replace each $e_j$ by $e_j'$ on the network, that is, add $e_1',e_2',\cdots,e_l'$ on the network and delete $e_1,e_2,\cdots,e_l$ from the network. Then the rank of the error pattern $\p$ with respect to a sink node $t$ in the original network is equal to the minimum cut capacity between $s_{\p}$ and $t$.
\end{prop}

\begin{defn}[{\cite[Definition 6]{zhang-correction}}]
An $\w$-dimensional linear network error correction code is called a regular code if for any $t\in T$, $\dim(\Phi(t))=\w$, or equivalently, $\Rank(F_t)=\w$.
\end{defn}

If the considered code is not regular, i.e., $\Rank(F_t)<\w$ for at least one sink node $t\in T$, then even in the error-free case, the code is not decodable at at least one sink node $t\in T$, not to mention network error correction. Therefore, we must consider regular codes for all information rates.

\begin{defn}[{\cite[Definition 7]{zhang-correction}}]
The minimum distance of a regular linear network error correction code at sink node $t$ is defined as
$$d_{\min}^{(t)}=\min\{ rank_t(\p):\ \dim(\Delta(t,\p)\cap \Phi(t))>0 \}.$$
\end{defn}

Now, for linear network error correction codes, we give the refined Singleton bound as follows.
\begin{prop}[The Refined Singleton Bound]\label{thm_ref_singleton_b}
Let $d_{\min}^{(t)}$ be the minimum distance of a regular linear network error correction code at a sink node $t\in T$. Then
$$d_{\min}^{(t)}\leq \delta_t+1,$$
where $\delta_t=C_t-\w$ is called the redundancy of the sink node $t$ with $C_t$ being the minimum cut capacity between $s$ and $t$, and $\w$ being the information rate.
\end{prop}

We adopt the convention that the regular linear network error correction codes satisfying the refined Singleton bound with equality for all sink nodes are called linear network error correction maximum distance separable (MDS) codes, or network MDS codes for short.

\section{Variable-Rate Network Error Correction MDS Codes}

In a single source finite acyclic communication network $G$, assume that the source transmits the messages at several distinct rates $\w_1,\w_2,\cdots,\w_h$ within a session, and let $\w=\max\{ \w_1,\w_2,\cdots,\w_h \}$ satisfying $\w\leq \min_{t\in T}C_t$ to avoid triviality, where again $C_t$ is the minimum cut capacity between the source node $s$ and the sink node $t$.

According to the constructive algorithm of linear network error correction codes \cite[Algorithm 1]{Guang-MDS}, we know that an $\w$-dimensional linear network error correction MDS code can be designed on $G$. In this section, we will show that if one $\w$-dimensional network MDS code is given, then an $(\w-1)$-dimensional network MDS code with the same local encoding kernels at all non-source nodes can also be constructed. Then a constructive algorithm is proposed. By using this algorithm recursively, we can construct all $\w_i$-dimensional $(1\leq i\leq h)$ linear network MDS codes with the same local encoding kernels at all non-source nodes.

First, we need several lemmas as follows.

\begin{lemma}\label{lem_1}
Let $\{\f_e:\ e\in E  \}$ constitute a global description of a regular linear network error correction code over a network $G$,
and $\vk=[k_1\ k_2\ \cdots\ k_{\w-1}]^{\top}\in \mF^{\w-1}$ be an arbitrary $(\w-1)$-dimensional column vector.
Define the matrix
$$F_t^{(\w-1)}(\vk)=\begin{bmatrix}I_{\w-1} & \vk\end{bmatrix}\cdot F_t,$$
where $I_{\w-1}$ is an $(\w-1)\times(\w-1)$ identity matrix. Then the row vectors of $F_t^{(\w-1)}(\vk)$ are still linearly independent, i.e., $\Rank(F_t^{(\w-1)}(\vk))=\w-1$.
\end{lemma}
\begin{IEEEproof}
For each sink node $t\in T$, we know
$$F_t=\begin{bmatrix}
\rt(d'_1)\\
\vdots\\
\rt(d'_\w)
\end{bmatrix}.$$
Consequently,
\begin{align*}
F_t^{(\w-1)}(\vk)&=\begin{bmatrix}I_{\w-1} & \vk\end{bmatrix}\cdot F_t\\
&=\begin{bmatrix}
\rt(d'_1)+k_1\cdot\rt(d'_\w)\\
\rt(d'_2)+k_2\cdot\rt(d'_\w)\\
\cdots\cdots\\
\rt(d'_{\w-1})+k_{\w-1}\cdot\rt(d'_\w)
\end{bmatrix}.
\end{align*}

To simply notation, let $r_i=\rt(d'_i)$ for all $1\leq i \leq \w$.
It follows that we only need to prove that, for any $(\w-1)$-dimensional vector
$\vk=[k_1\ k_2\ \cdots\ k_{\w-1}]^{\top}\in \mF^{\w-1}$, the $(\w-1)$ row vectors $r_1'\triangleq r_1+k_1r_\w$, $r_2'\triangleq r_2+k_2r_\w$, $\cdots$, $r_{\w-1}'\triangleq r_{\w-1}+k_{\w-1}r_\w$ are linearly independent. Conversely, suppose that $r_1',r_2',\cdots,r_{\w-1}'$ are linearly dependent. This implies that there exist $(\w-1)$ elements
$a_1,a_2,\cdots,a_{\w-1}$ of $\mF$, not all $0$, such that
\begin{equation*}\label{eq_com}
a_{1}r_{1}'+a_{2}r_{2}'+\cdots+a_{\w-1}r_{\w-1}'=0,
\end{equation*}
that is,
\begin{equation*}
a_{1}r_{1}+a_{2}r_{2}+\cdots+a_{\w-1}r_{\w-1}+(a_{1}k_{1}+a_{2}k_{2}+\cdots+a_{\w-1}k_{\w-1})r_{\w}=0.
\end{equation*}
Since $r_{1},r_{2},\cdots,r_{\w}$ are linearly independent vectors, we further have
$$a_{1}=a_{2}=\cdots=a_{\w-1}=a_{1}k_{1}+a_{2}k_{2}+\cdots+a_{\w-1}k_{\w-1}=0,$$
particularly,
$$a_{1}=a_{2}=\cdots=a_{\w-1}=0,$$
which is a contradiction. So $r_1',r_2',\cdots,r_{\w-1}'$ are linearly independent.
That is, for any $(\w-1)$-dimensional column vector $\vk=[k_1\ k_2\ \cdots\ k_{\w-1}]^{\top}\in \mF^{\w-1}$, one has
$$\Rank(F_{t}^{\w-1}(\vk))=\Rank\big(\begin{bmatrix}r_1'^{\top}&\cdots&r_{\w-1}'^{\top}\end{bmatrix}^{\top}\big)=\w-1.$$
The lemma is proved.
\end{IEEEproof}

Let $\mathbf{C}_\w$ be an $\w$-dimensional $\mF$-valued regular linear network error correction code over an acyclic network $G=(V,E)$, and $\f_e$ represent the extended global encoding
kernel of the channel $e$ for all $e\in E$. Let $I_{\w-1}$ and $I_{\mE}$ denote the
$(\w-1)\times(\w-1)$ and $\mE\times \mE$ identity matrices, respectively. Let $\vk=[k_1\ k_2\ \cdots\ k_{\w-1}]^{\top}\in \mF^{\w-1}$ be an arbitrary $(\w-1)$-dimensional column vector. For each non-imaginary channel $e$, define
\begin{align}\label{ext_f_w-1}
\f_{e}^{(\w-1)}(\vk)=\begin{bmatrix} I_{\w-1}& \vk & \bzero_{(\w-1)\times\mE}\\
\bzero_{\mE\times(\w-1)}&\bzero_{\mE\times1}&I_{\mE}\end{bmatrix}\cdot\f_{e},
\end{align}
where $\bzero_{a\times b}$ represents the $a\times b$ all-zero matrix.

\begin{lemma}\label{lem_2}
If $\{ \f_{e}:e\in E \}$ constitutes a global description of an $\w$-dimensional $\mF$-valued regular linear network error correction code $\mathbf{C}_\w$ over an acyclic network $G$, then $\{ \f_{e}^{(\w-1)}(\vk):e\in E \}$ constitutes a global description of an $(\w-1)$-dimensional regular linear network
error correction code for the network $G$. In particular, the local encoding kernel of this $(\w-1)$-dimensional code at each non-source node is the same as that of the original $\w$-dimensional code $\mathbf{C}_\w$.
\end{lemma}
\begin{IEEEproof}
Let $k_{d,e}\in \mF$ be the local encoding coefficient of the original $\w$-dimensional code $\mathbf{C}_\w$ for the adjacent pair $(d,e)$ of channels. First, we show that
$\{ \f_{e}^{(\w-1)}(\vk):e\in E \}$ constitutes an $(\w-1)$-dimensional linear network error correction code by demonstrating the existence of the corresponding local encoding coefficient $k_{d,e}^{(\w-1)}$ for the adjacent channel pair $(d,e)$ of channels, where $\vk=[k_1\ k_2\ \cdots\ k_{\w-1}]^{\top}\in \mF^{\w-1}$.

By convention, assume that the extended global encoding kernels of the $(\w-1)$ imaginary message channels are
\begin{align*}
\f_{d'_1}^{(\w-1)}=\begin{bmatrix}f_{d'_1}^{(\w-1)}\\\bzero_{\mE\times1} \end{bmatrix},\ \f_{d'_2}^{(\w-1)}=\begin{bmatrix}f_{d'_2}^{(\w-1)}\\\bzero_{\mE\times1} \end{bmatrix},\ \cdots,\
\f_{d'_{\w-1}}^{(\w-1)}=\begin{bmatrix}f_{d'_{\w-1}}^{(\w-1)}\\\bzero_{\mE\times1} \end{bmatrix},
\end{align*}
where $f_{d'_1}^{(\w-1)},f_{d'_2}^{(\w-1)},\cdots,f_{d'_{\w-1}}^{(\w-1)}$ form the standard basis of $\mF^{\w-1}$.

\textit{\textbf{Case 1.}} For each channel $e\in Out(s)$, we have
\begin{align}\label{equ_f}
\f_{e}^{(\w-1)}(\vk)&=\begin{bmatrix} I_{\w-1}& \vk & \bzero_{(\w-1)\times\mE}\\
\bzero_{\mE\times(\w-1)}&\bzero_{\mE\times1}&I_{\mE}\end{bmatrix}\cdot\f_{e}\nonumber\\
&=\begin{bmatrix} I_{\w-1}& \vk & \bzero_{(\w-1)\times\mE}\\
\bzero_{\mE\times(\w-1)}&\bzero_{\mE\times1}&I_{\mE}\end{bmatrix}\cdot\begin{bmatrix}f_e\\g_e\end{bmatrix}\nonumber\\
&=\begin{bmatrix}f_e(d'_1)+k_{1}f_{e}(d'_{\w})\\f_{e}(d'_{2})+k_{2}f_{e}(d'_{\w})\\
\cdots\cdots\cdots\\f_e(d'_{\w-1})+k_{\w-1}f_{e}(d'_{\w})\\g_e\end{bmatrix}=\begin{bmatrix}k_{d'_1,e}+k_{1}k_{d'_\w,e}\\k_{d'_2,e}+k_{2}k_{d'_\w,e}\\
\cdots\cdots\cdots\\k_{d'_{\w-1},e}+k_{\w-1}k_{d'_\w,e}\\g_e
\end{bmatrix},
\end{align}
where the equation (\ref{equ_f}) follows from $f_e(d'_i)=k_{d'_i,e}$, the local encoding coefficient for the adjacent pair $(d_i', e)$, $1\leq i \leq \w$. Further, define $k_{d'_i,e}^{(\w-1)}(\vk)=k_{d'_i,e}+k_{i}k_{d'_\w,e}$, $i=1,2,\cdots,\w-1$. Thus
\begin{align*}
\f_e^{(\w-1)}(\vk)=&\sum_{d\in In(s)}k_{d,e}^{(\w-1)}(\vk)\cdot\f_d^{(\w-1)}+1_e^{(\w-1)}\\
=&\sum_{i=1}^{\w-1}k_{d'_i,e}^{(\w-1)}(\vk)\cdot\f_{d'_i}^{(\w-1)}+1_e^{(\w-1)},
\end{align*}
where $1_e^{(\w-1)}$ is an $((\w-1)+\mE)$-dimensional column vector which is the indicator function of the channel $e$.

\textit{\textbf{Case 2.}} For other non-imaginary channels $e\notin Out(s)$, we know from (\ref{equ_ext_f})
$$\f_e=\sum_{d\in In(tail(e))}k_{d,e}\cdot\f_d+1_e.$$
Multiplying both sides by
$\begin{bmatrix} I_{\w-1}& \vk & \bzero_{(\w-1)\times\mE}\\
\bzero_{\mE\times(\w-1)}&\bzero_{\mE\times1}&I_{\mE}\end{bmatrix}$, together with (\ref{ext_f_w-1}),
yields that
\begin{align*}
\f_e^{(\w-1)}(\vk)=&\begin{bmatrix} I_{\w-1}& \vk & \bzero_{(\w-1)\times\mE}\\
\bzero_{\mE\times(\w-1)}&\bzero_{\mE\times1}&I_{\mE}\end{bmatrix}\cdot\f_e\\
=&\sum_{d\in In(tail(e))}k_{d,e}\begin{bmatrix} I_{\w-1}& \vk & \bzero_{(\w-1)\times\mE}\\
\bzero_{\mE\times(\w-1)}&\bzero_{\mE\times1}&I_{\mE}\end{bmatrix}\cdot\f_d+\begin{bmatrix} I_{\w-1}& \vk & \bzero_{(\w-1)\times\mE}\\
\bzero_{\mE\times(\w-1)}&\bzero_{\mE\times1}&I_{\mE}\end{bmatrix}\cdot1_e\\
=&\sum_{d\in In(tail(e))}k_{d,e}\cdot\f_d^{(\w-1)}(\vk)+1_e^{(\w-1)},
\end{align*}
which leads to $k_{d,e}^{(\w-1)}(\vk)=k_{d,e}$ for all adjacent pairs $(d,e)$ of channels $d,e\in E$.
Combining the above two cases, $\{ f_e^{(\w-1)}(\vk):e\in E \}$ consists of all the extended global encoding kernels of an $(\w-1)$-dimensional linear network error correction code, and for each adjacent pair $(d,e)$ of channels $d,e\in E$, $k_{d,e}$ is also the local encoding coefficient of this
$(\w-1)$-dimensional code.

Applying Lemma \ref{lem_1} and the fact that the $\w$-dimensional linear network error correction code $\mathbf{C}_\w$ is
regular, $\{ \f_e^{(\w-1)}(\vk):e\in E \}$ also constitutes an $(\w-1)$-dimensional regular linear network error correction code. This completes the proof.
\end{IEEEproof}

Moreover, we need the following lemma, which gives three equivalent relations on the minimum distance.
\begin{lemma}[{\cite[Proposition 2]{Guang-MDS}}]\label{lem_3}
For the minimum distance of a regular linear network error correction code at every sink node $t$,
we have the following equalities:
\begin{align*}
d_{\min}^{(t)}&=\min\{ rank_t(\p):\ \Delta(t,\p)\cap \Phi(t)\neq \{\bzero\} \}\\
              &=\min\{ |\p|:\ \Delta(t,\p)\cap \Phi(t)\neq \{\bzero\} \}\\
              &=\min\{ \dim(\Delta(t,\p)):\ \Delta(t,\p)\cap \Phi(t)\neq \{\bzero\} \}.
\end{align*}
\end{lemma}

For a network MDS code, define a set of error patterns for each sink node $t\in T$:
$$
Q(t)=\Big\{\mbox{error\;pattern\;}\p: \Delta(t,\p)\cap \Phi(t)\neq \{\bzero\}\mbox{ and } |\p|=d_{\min}^{(t)} \Big\}.
$$

\begin{lemma}\label{lem_4}
For an $\w$-dimensional network MDS code on $G$, we have for any sink node $t\in T$ and any error pattern $\p\in Q(t)$,
$$\dim(\Delta(t,\p)\cap \Phi(t))=1.$$
\end{lemma}
\begin{IEEEproof}
For the given network MDS code, we know $d_{\min}^{(t)}=C_t-\w+1=\dt+1$ for each sink node $t\in T$. This means that
$$Q(t)=\Big\{\mbox{error\;pattern\;} \p: \Delta(t,\p)\cap \Phi(t)\neq \{\bzero\} \mbox{ and } |\p|=\dt+1 \Big\}.$$
For any $\p\in Q(t)$, $|\p|=\dt+1$ implies that
\begin{equation}\label{eq_4}
\dim(\Delta(t,\p))\leq |\p|=\dt+1 .
\end{equation}
On the other hand, by Lemma \ref{lem_3} and the definition of network MDS codes, it is readily seen that
$$d_{\min}^{(t)}=\min\{ \dim(\Delta(t,\p')):  \Delta(t,\p')\cap \Phi(t)\neq \{\bzero\}\}=\dt+1.$$
Together with $\Delta(t,\p)\cap \Phi(t)\neq \{\bzero\}$, it follows that
\begin{equation}\label{eq_5}
\dim(\Delta(t,\p))\geq  d_{\min}^{(t)}=\dt+1 .
\end{equation}
Combining the inequalities (\ref{eq_4}) and (\ref{eq_5}), one has
\begin{align}\label{eq_5.1}
\dim(\Delta(t,\p))=\dt+1=|\p|.
\end{align}

For simplicity, let $d=d_{\min}^{(t)}=\dt+1$, $\p=\{ e_1,e_2,\cdots, e_d\}$ and $r_i\triangleq \rt(e_i)$, $i=1,2,\cdots,d$. Hence, $r_1,r_2,\cdots,r_d$ are $d$ linearly independent vectors since $\dim(\Delta(t,\p))=|\p|=d$ from (\ref{eq_5.1}) and $\Delta(t,\p)=\langle\{ r_i:\ 1\leq i \leq d \}\rangle$.

Suppose that $\dim(\Delta(t,\p)\cap \Phi(t))\geq 2$ and then let $\vec{l_1},\ \vec{l_2}$ be two linearly independent vectors in the vector space $\Delta(t,\p)\cap \Phi(t)$. Then there exist $a_1,a_2,\cdots,a_d$ in $\mF$, not all $0$, and $b_1,b_2,\cdots,b_d$ in $\mF$, not all $0$, such that
$$\begin{cases}
\vec{l_1}=a_1r_1+a_2r_2+\cdots+a_dr_d\\
\vec{l_2}=b_1r_1+b_2r_2+\cdots+b_dr_d.
\end{cases}$$
Further, for all $i=1,2,\cdots,d$, we claim that either $a_i$ or $b_i$ is zero.
Assume the contrary, that is, there exists some $i\ (1\leq i \leq d)$ such that $a_i\neq 0$, $b_i\neq 0$.
If so, we have
$$a_i\vec{l_2}-b_i\vec{l_1}\in \Delta(t,\p\backslash \{e_i\})\cap \Phi(t)$$ and $$a_i\vec{l_2}-b_i\vec{l_1}\neq \bzero$$ because of
the linear independence between $\vec{l_1}$ and $\vec{l_2}$, which means that
$\Delta(t,\p\backslash \{e_i\})\cap\Phi(t)\neq \{\bzero\}$.
Hence,
\begin{align*}
d_{\min}^{(t)}=&\min\{ \dim(\Delta(t,\p')):  \Delta(t,\p')\cap \Phi(t)\neq \{\bzero\}\}\\
\leq &\dim(\Delta(t,\p\backslash \{e_i\}))=\dt,
\end{align*}
which is a contradiction to $d_{\min}^{(t)}=\dt+1$. Now, we can say that for all $i=1,2,\cdots,d$, either $a_i=0$ or $b_i=0$.

Without loss of generality, assume $a_1\neq 0$ and $b_1=0$.
That is, the non-zero vector
\begin{align*}
\vec{l_2}&=b_1r_1+b_2r_2+\cdots+b_dr_d\\
         &=b_2r_2+\cdots+b_dr_d\in \Delta(t,\p\backslash \{e_1\}),
\end{align*}
which, together with $\vec{l_2}\in \Delta(t,\p)\cap\Phi(t)$, leads to
$$\bzero\neq \vec{l_2}\in \Phi(t)\cap\Delta(t,\p\backslash \{e_1\}).$$
It also follows that
\begin{align*}
d_{\min}^{(t)}&=\min\{ \dim(\Delta(t,\p')):\Phi(t)\cap\Delta(t,\p')\neq \{\bzero\}\}\\
&\leq \dim(\Delta(t,\p\backslash \{e_1\}))=\dt.
\end{align*}
This also violates the condition $d_{\min}^{(t)}=\dt+1$.

Therefore, we have shown that $\dim(\Delta(t,\p)\cap \Phi(t))=1$ for any $\p\in Q(t)$. This completes the proof.
\end{IEEEproof}

\begin{lemma}\label{lem_5}
For an acyclic network $G$, an $\w$-dimensional $\mF$-valued linear network MDS code with field size $|\mF|>\sum_{t\in T}|Q(t)|$ is given. Then there exists an $(\w-1)$-dimensional column vector $\vk=[k_1\ k_2\ \cdots\ k_{\w-1}]^{\top} \in \mF^{\w-1}$ such that
$$\dim(\Delta(t,\p)\cap \Phi^{(\w-1)}(t,\vk))=0$$
for each sink node $t\in T$ and each error pattern $\p\in Q(t)$, where
$$\Phi^{(\w-1)}(t,\vk)=\langle \{\rt(d'_i)+k_i\cdot\rt(d'_\w): 1\leq i \leq \w-1 \}\rangle.$$
\end{lemma}
\begin{IEEEproof}
First, we show that, when a fixed sink node $t\in T$ and a fixed $\p\in Q(t)$ are under consideration, there exists an $(\w-1)$-dimensional column vector $\vk\in \mF^{\w-1}$ such that $\Delta(t,\p)\cap \Phi^{(\w-1)}(t,\vk)=\{\bzero\}$.
Conversely, suppose that for any $\vk=[k_1\ k_2\ \cdots\ k_{\w-1}]^{\top}\in \mF^{\w-1}$,
\begin{equation}\label{eq_6}
\Delta(t,\p)\cap \Phi^{(\w-1)}(t,\vk)\neq \{\bzero\}.
\end{equation}
Clearly, $\Phi^{(\w-1)}(t,\vk)\subseteq \Phi(t)$, which shows that
\begin{equation}\label{eq_7}
\Delta(t,\p)\cap \Phi^{(\w-1)}(t,\vk)\subseteq \Delta(t,\p)\cap \Phi(t).
\end{equation}
Using formulae (\ref{eq_6}), (\ref{eq_7}) and $\dim(\Delta(t,\p)\cap \Phi(t))=1$ from Lemma \ref{lem_4}, we have $$\Delta(t,\p)\cap \Phi^{(\w-1)}(t,\vk)=\Delta(t,\p)\cap \Phi(t).$$

To simply notation, again let $r_i=\rt(d'_i)$, $1\leq i \leq \w$ and $r_j'=r_j+k_jr_\w$, $1\leq j \leq \w-1$. Then $r_1,r_2,\cdots,r_\w$ form a basis of vector space $\Phi(t)$, and $r_1',r_2',\cdots,r_{\w-1}'$ form a basis of vector space $\Phi^{(\w-1)}(t,\vk)$ since $r_1',r_2',\cdots,r_{\w-1}'$ are linearly independent from Lemma \ref{lem_1}.

Let $\vec{l}$ be a non-zero vector in $\Delta(t,\p)\cap \Phi(t)$. Then there exist unique elements $a_1,a_2,\cdots,a_\w \in \mF$, not all $0$, such that
\begin{align}\label{eq_8}
\vec{l}=a_1r_1+a_2r_2+\cdots+a_{\w-1}r_{\w-1}+a_{\w}r_\w.
\end{align}
Moreover, it is certain that $\vec{l}\in \Delta(t,\p)\cap \Phi^{(\w-1)}(t,\vk)$. This means that there also exist unique elements $b_1,b_2,\cdots,b_{\w-1}\in \mF$ such that
\begin{align*}
\vec{l}=b_1r_1'+b_2r_2'+\cdots+b_{\w-1}r_{\w-1}'.
\end{align*}
Hence,
\begin{align}\label{eq_9}
\vec{l}=&b_1(r_1+k_1r_\w)+b_2(r_2+k_2r_\w)+\cdots+b_{\w-1}(r_{\w-1}+k_{\w-1}r_\w)\nonumber \\
=&b_1r_1+b_2r_2+\cdots+b_{\w-1}r_{\w-1}+(b_1k_1+b_2k_2+\cdots+b_{\w-1}k_{\w-1})r_\w.
\end{align}
Due to both representations (\ref{eq_8}) and (\ref{eq_9}) of $\vec{l}$, one has $a_i=b_i$ for $1\leq i\leq \w-1$ and \begin{align*}
a_\w&=b_1k_1+b_2k_2+\cdots+b_{\w-1}k_{\w-1}\\
   &=a_1k_1+a_2k_2+\cdots+a_{\w-1}k_{\w-1}.
\end{align*}
This implies that, for any $\vk=[k_1\ k_2\ \cdots\ k_{\w-1}]^{\top}\in \mF^{\w-1}$, it always follows
$$a_\w=a_1k_1+a_2k_2+\cdots+a_{\w-1}k_{\w-1},$$
which is obviously impossible. Therefore, there exists an $(\w-1)$-dimensional column vector $\vk\in \mF^{\w-1}$ such that $\Delta(t,\p)\cap \Phi^{(\w-1)}(t,\vk)=\{\bzero\}$.

Furthermore, consider the following set:
\begin{align*}
K(t,\p)=\Big\{ \vk=[k_1\ k_2\ \cdots\ k_{\w-1}]^{\top}\in \mF^{\w-1}: \sum_{i=1}^{\w-1}a_ik_i=a_\w \Big\}.
\end{align*}
It is not hard to see $|K(t,\p)|=|\mF|^{\w-2}$, and, because
$0\leq \dim(\Delta(t,\p)\cap \Phi^{(\w-1)}(t,\vk))\leq 1$,
$$\dim(\Delta(t,\p)\cap \Phi^{(\w-1)}(t,\vk))=0$$
for any $\vk\in \mF^{\w-1}\backslash K(t,\p)$. Thus, $K(t,\p)$ can be rewritten as the following equivalent form:
\begin{align*}
K(t,\p)=
\Big\{ \vk\in \mF^{\w-1}: \Delta(t,\p)\cap \Phi(t)=\Delta(t,\p)\cap \Phi^{(\w-1)}(t,\vk) \Big\}.
\end{align*}
As a result, for any
$\vk\in \mF^{\w-1}\backslash \cup_{t\in T}\cup_{\p\in Q(t)}K(t,\p)$, we have $\Delta(t,\p)\cap \Phi^{(\w-1)}(t,\vk)=\{\bzero\}$ for any $t\in T$ and $\p\in Q(t)$.

At last, we show that $|\mF^{\w-1}\backslash \cup_{t\in T}\cup_{\p\in Q(t)}K(t,\p)|>0$ under the condition $|\mF|>\sum_{t\in T}|Q(t)|$. This follows because
\begin{equation*}
\begin{split}
&\big|\mF^{\w-1}\backslash \cup_{t\in T}\cup_{\p\in Q(t)}K(t,\p)\big|\\
=&\big|\mF^{\w-1}\big|-\big|\mF^{\w-1}\cap [\cup_{t\in T}\cup_{\p\in Q(t)}K(t,\p)]\big|\\
=&\big|\mF\big|^{\w-1}-\big|\cup_{t\in T}\cup_{\p\in Q(t)}\mF^{\w-1}\cap K(t,\p)\big|\\
\geq&\big|\mF\big|^{\w-1}-\sum_{t\in T}\sum_{\p\in Q(t)}\big|\mF^{\w-1}\cap K(t,\p)\big|\\
=&\big|\mF\big|^{\w-1}-\big|\mF\big|^{\w-2}\sum_{t\in T}\big|Q(t)\big|\\
=&\big|\mF\big|^{\w-2}\big[\big|\mF\big|-\sum_{t\in T}\big|Q(t)\big|\big]\\
>&0.
\end{split}
\end{equation*}
The lemma is proved.
\end{IEEEproof}

Under the support of the above five lemmas, we can give the main theorem below.

\begin{thm}\label{thm_w-1}
Let $\mathbf{C}_\w$ be an $\w$-dimensional $\mF$-valued network MDS code. If the size of the base field $\mF$ satisfies $|\mF|>\sum_{t\in T}|Q(t)|$, then there exists an $(\w-1)$-dimensional $\mF$-valued network MDS code for this network $G$ with the same local encoding kernels at all non-source nodes as that of $\mathbf{C}_\w$.
\end{thm}
\begin{IEEEproof}
For the given network MDS code on an acyclic network $G$, Lemmas \ref{lem_2} and \ref{lem_5} imply that there exists an $(\w-1)$-dimensional column vector $\vk\in \mF^{\w-1}$ such that $\{\f_{e}^{(\w-1)}(\vk):\ e\in E\}$ is the set of all extended global encoding kernels of an $(\w-1)$-dimensional regular linear network error correction code, and
$$\Delta(t,\p)\cap \Phi^{(\w-1)}(t,\vk)=\{\bzero\}$$
for any $t\in T$ and any error pattern $\p\in Q(t)$.

On the other hand, by the definition of network MDS codes and Lemma \ref{lem_3}, we know
$$\Delta(t,\p)\cap \Phi(t)=\{\bzero\}$$
for all error patterns $\p$ with $|\p|<\dt+1$. Hence, for any error pattern $\p$ satisfying $|\p|<\dt+1$, or $|\p|=\dt+1$ but $\Delta(t,\p)\cap \Phi(t)=\{\bzero\}$, one has that $\Delta(t,\p)\cap \Phi^{(\w-1)}(t,\vk)=\{\bzero\}$ since $\Phi^{(\w-1)}(t,\vk)\subseteq \Phi(t)$.

Combining the above, for any $t\in T$ and any error pattern $\p$ with $|\p|\leq \dt+1$, it always follows
$$\Delta(t,\p)\cap \Phi^{(\w-1)}(t,\vk)=\{\bzero\},$$
which implies that
$$
d_{\min}^{(t,\w-1)}\triangleq \min\{ |\p|:  \Delta(t,\p)\cap \Phi^{(\w-1)}(t,\vk)\neq\{\bzero\} \}\geq \dt+2=C_t-(\w-1)+1.
$$
On the other hand, the refined Singleton bound on linear network error correction codes (Proposition \ref{thm_ref_singleton_b}) indicates that, for each sink node $t\in T$,
$$
d_{\min}^{(t,\w-1)}\leq C_t-(\w-1)+1=\dt+2.
$$
Thus, $d_{\min}^{(t,\w-1)}=C_t-(\w-1)+1=\dt+2$. That is, $\{\f_{e}^{(\w-1)}(\vk):e\in E\}$ constitutes a global description of an $(\w-1)$-dimensional network MDS code on the network $G$, which completes the proof.
\end{IEEEproof}

Again let $\mathbf{C}_\w$ be an $\w$-dimensional $\mF$-valued network MDS code over an acyclic network $G$. Using the above constructive method recursively, if the field size $|\mF|$ is big enough, then, for any information rate $\w'\leq \w$, it is feasible to construct an $\w'$-dimensional $\mF$-valued network MDS code over the network $G$ satisfying the condition that the local encoding kernels of this $\w'$-dimensional network MDS code at all internal nodes are the same as that of the original $\w$-dimensional network MDS code. These network MDS codes with the same local encoding kernels at all internal nodes are called \textit{a family of variable-rate network MDS codes}.

By \cite[Theorem 5 and Algorithm 1]{Guang-MDS}, it follows that if $|\mF|\geq \sum_{t\in T}|R_t(\dt)|$ where
$$R_t(\dt)=\{\mbox{error pattern}\ \p:\ |\p|=rank_t(\p)=\dt\}$$
and $\dt=C_t-\w$, we can construct an $\w$-dimensional network MDS code $\mathbf{C}_\w$. By Theorem \ref{thm_w-1}, if $|\mF|>\sum_{t\in T}|Q(t)|$, where recall that
$$
Q(t)=\Big\{\mbox{error\;pattern\;}\p: \Delta(t,\p)\cap \Phi(t)\neq \{\bzero\}\mbox{ and } |\p|
=d_{\min}^{(t)} \Big\},
$$
we can construct an $(\w-1)$-dimensional network MDS code $\mathbf{C}_{\w-1}$ with the same local encoding kernels at all internal nodes as that of $\mathbf{C}_\w$. Subsequently, for any error pattern $\p$ with $rank_t(\p)<\dt+1$, we have $\Delta(t,\p)\cap\Phi(t)=\{\bzero\}$ because of the definition of the minimum distance
$$d_{\min}^{(t)}\triangleq \min\{ rank_t(\p):\ \Delta(t,\p)\cap\Phi(t)\neq\{\bzero\} \}$$
and the refined Singleton bound (Proposition \ref{thm_ref_singleton_b}) that $d_{\min}^{(t)}\leq C_t-\w+1=\dt+1$. Consequently, we have
\begin{align*}
Q(t)&=\Big\{\p\subseteq E: \Delta(t,\p)\cap \Phi(t)\neq \{\bzero\}\mbox{ and } rank_t(\p)=|\p|
=\dt+1 \Big\}\\
&\subseteq \Big\{\p\subseteq E: rank_t(\p)=|\p|
=\dt+1\Big\}\\
&=R_t(\dt+1)\triangleq R_t(\dt^{(\w-1)}),
\end{align*}
where $\dt^{(\w-1)}=C_t-(\w-1)=\dt+1$.
Therefore, if the base field size
$$|\mF|>\max\Big\{ \sum_{t\in T}|R_t(\dt)|,\ \sum_{t\in T}|R_t(\dt+1)| \Big\},$$
then, applying our approach, we can construct two variable-rate network MDS codes with respective information rates $\w$ and $\w-1$, that is, the constructed $(\w-1)$-dimensional and $\w$-dimensional network MDS codes $\mC_{\w-1}$ and $\mC_{\w}$ have the same local encoding kernels at all internal nodes.

Recursively, if the field size satisfies
$$|\mF|>\max\Big\{ \sum_{t\in T}|R_t(\dt)|,\sum_{t\in T}|R_t(\dt+1)|,\cdots, \sum_{t\in T}|R_t(\dt+\w-1)| \Big\},$$
or equivalently,
$$|\mF|>\max\Big\{ \sum_{t\in T}|R_t(C_t-\w)|,\sum_{t\in T}|R_t(C_t-(\w-1))|,\cdots, \sum_{t\in T}|R_t(C_t-1)| \Big\},$$
we can construct all $\w'$-dimensional $(1\leq \w'\leq \w)$ network MDS codes having the same local encoding kernel at each internal node. Therefore, we have the following theorem.
\begin{thm}
For a single source multicast acyclic network $G$, if the size of the base field satisfies
$$|\mF|>\max_{0\leq i \leq \w-1 }\sum_{t\in T}\big|R_t(\dt+i)\big|=\max_{0\leq i \leq \w-1 }\sum_{t\in T}\big|R_t(C_t-\w+i)\big|,$$
then we can construct a family of variable-rate $\mF$-valued network MDS codes of dimensions $1,2,\cdots,\w$.
\end{thm}

Further, we have for each $\w'$, $\sum_{t\in T}\big|R_t(C_t-\w')\big|\leq \sum_{t\in T}{|E| \choose C_t-\w'}$ from \cite[Lemma 6]{Guang-MDS}. Thus,
if
$$|\mF|>\max\Big\{ \sum_{t\in T}{|E| \choose \dt},\sum_{t\in T}{|E| \choose \dt+1},\cdots,\sum_{t\in T}{|E| \choose C_t-1}\Big\},$$
we are more able to construct a family of variable-rate network MDS codes of dimensions $1,2,\cdots, \w$, which have the same local encoding kernel at each internal node. This result can be described by the following corollary.

\begin{cor}\label{cor-8}
For a single source multicast network $G$, if the size of the base field satisfies $|\mF|>\max_{1\leq i\leq \w}\sum_{t\in T}{|E| \choose C_t-i}$, then we can construct a family of variable-rate $\mF$-valued network MDS codes of dimensions $1,2,\cdots,\w$.
\end{cor}
\begin{rem}
Generally speaking, in most communication networks, $C_t\leq \lfloor\frac{|E|}{2}\rfloor$ for any sink node $t\in T$. Therefore, $\max_{1\leq i\leq \w} \sum_{t\in T}{|E| \choose C_t-i} =\sum_{t\in T}{|E| \choose C_t-1}$. This shows that we can construct a family of variable-rate network MDS codes provided $|\mF|>\sum_{t\in T}{|E| \choose C_t-1}$.
\end{rem}

Now, we can give an algorithm for constructing a family of variable-rate network MDS codes based on our discussion above.

\begin{description}
  \item[Step 1: ] \ Construct an $\w$-dimensional network MDS code $\mathbf{C}_\w$ by Algorithm 1 in \cite{Guang-MDS};
  \item[Step 2: ] \ Choose an $(\w-1)$-dimensional column vector $\vk=[k_1\ k_2\ \cdots\ k_{\w-1}]^{\top}\in \mF^{\w-1}$ such that
  \begin{align}\label{choose-vk}
  \vk\in \mF^{\w-1}\backslash\cup_{t\in T}\cup_{\p\in Q(t)}K(t,\p),
  \end{align}
  where $K(t,\p)$ is a collection of $(\w-1)$-dimensional $\mF$-valued column vectors as defined in Lemma \ref{lem_5}.
  \item[Step 3: ] \ $\{ \f^{(\w-1)}_e(\vk): e\in E \}$ constitutes an $(\w-1)$-dimensional $\mF$-valued network MDS code with the same local encoding kernels at all internal nodes as that of $\mathbf{C}_\w$.
\end{description}

Using this algorithm recursively, we can construct a family of variable-rate network MDS codes of dimensions $1,2,\cdots,\w$.

\begin{rem}
For the proposed variable-rate network error correction problem, we have to simultaneously consider the information transmission and network error correction, or equivalently, the regular property and MDS property of the codes. If we assume that all channels are error-free, that is, only information transmission is under the consideration, our constructive algorithm degenerates into an algorithm to construct variable-rate linear network codes presented in \cite{Fong-Yeung-variable-rate} since \cite[Lemma 1]{Fong-Yeung-variable-rate} can be regarded as a special case of Lemma 1 in the present paper. Further, together with other conditions such as Lemma 3 and a similar result, Lemma 5, in \cite{Fong-Yeung-variable-rate}, it will become the algorithm of Fong and Yeung for constructing variable-rate linear broadcast and static linear broadcast network codes.
\end{rem}

Now, we give a simple example to show how to construct an $(\w-1)$-dimensional network MDS code from an $\w$-dimensional one satisfying that both network MDS codes have the same local encoding kernels at all internal nodes by applying the above algorithm.

\begin{eg}

Let $G$ be a network with $C_{t_1}=C_{t_2}=3$ as showed by Fig. \ref{fig}, and let $\w=2$.
\begin{figure}[!htb]
\centering
\begin{tikzpicture}[->,>=stealth',shorten >=1pt,auto,node distance=2.5cm,
                    thick]
  \tikzstyle{every state}=[fill=none,draw=black,text=black,minimum size=7mm]
  \tikzstyle{node}=[circle,fill=none,draw=white, minimum size=0.1mm]
  \node[state]           (s)                      {$s$};
  \node[state]           (i)[below       of=s]    {$i$};
  \node[state]           (t1)[below left of=i]    {$t_1$};
  \node[state]           (t2)[below right of=i]   {$t_2$};
  \node[node]            (a)[above left of=s,xshift=10mm]{};
  \node[node]            (b)[above right of=s,xshift=-10mm]{};
\path (s)   edge[bend right=45]  node[swap]{$e_1$}(t1)
            edge                 node[swap]{$e_2$}(t1)
            edge                 node       {$e_3$}(i)
            edge[bend left=45]   node      {$e_5$}(t2)
            edge                 node      {$e_4$}(t2)
      (i)   edge              node         {$e_6$}(t1)
            edge              node[swap]{$e_7$}(t2)
      (a)   edge[dashed]      node[swap]{$d'_1$}(s)
      (b)   edge[dashed]      node{$d'_2$}(s);
\end{tikzpicture}
\caption{The network $G$ with $C_{t_1}=C_{t_2}=3$.}\label{fig}
\end{figure}
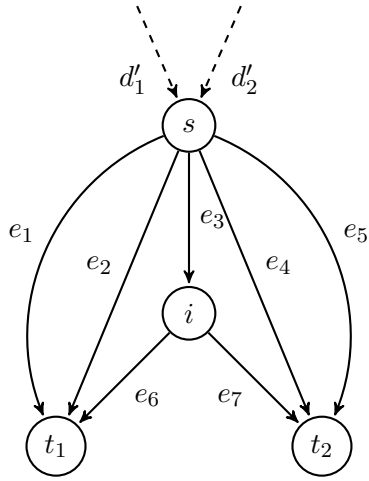
For simplicity, for all $d'_i \in In(s),\ e_j\in Out(s)$, denote by $k_{d'_i,j}$ the local encoding coefficient of the adjacent channel pair $(d'_i,e_j)$; and for $e_i,e_j\in E$ with $tail(e_j)=head(e_i)$, denote by $k_{i,j}$ the local encoding coefficient of the adjacent channel pair $(e_i,e_j)$.
Let the base field $\mF$ be $\mathbb{F}_3$, and let $$k_{d'_1,3}=k_{d'_2,2}=k_{d'_2,5}=0$$ and $$k_{d'_1,1}=k_{d'_1,2}=k_{d'_1,4}=k_{d'_1,5}=k_{d'_2,1}=k_{d'_2,3}=k_{d'_2,4}=k_{3,6}=k_{3,7}=1.$$
Then the extended global encoding kernels of all channels are
$$\f_{d'_1}=\begin{bmatrix}1&0&0&0&0&0&0&0&0\end{bmatrix}^\top,\qquad \f_{d'_2}=\begin{bmatrix}0&1&0&0&0&0&0&0&0\end{bmatrix}^\top,$$
$$\f_{e_1}=\begin{bmatrix}1&1&1&0&0&0&0&0&0\end{bmatrix}^\top,\qquad \f_{e_2}=\begin{bmatrix}1&0&0&1&0&0&0&0&0\end{bmatrix}^\top,$$
$$\f_{e_3}=\begin{bmatrix}0&1&0&0&1&0&0&0&0\end{bmatrix}^\top,\qquad \f_{e_4}=\begin{bmatrix}1&1&0&0&0&1&0&0&0\end{bmatrix}^\top,$$
$$\f_{e_5}=\begin{bmatrix}1&0&0&0&0&0&1&0&0\end{bmatrix}^\top,\qquad \f_{e_6}=\begin{bmatrix}0&1&0&0&1&0&0&1&0\end{bmatrix}^\top,$$
$$\f_{e_7}=\begin{bmatrix}0&1&0&0&1&0&0&0&1\end{bmatrix}^\top.$$
The decoding matrices at sink nodes $t_1$ and $t_2$ are given respectively by
$$\ti{F}_{t_1}=\begin{bmatrix}1&1&0\\1&0&1\\1&0&0\\0&1&0\\0&0&1\\0&0&0\\0&0&0\\0&0&1\\0&0&0
\end{bmatrix},\ \mbox{ and }
\ti{F}_{t_2}=\begin{bmatrix}1&1&0\\1&0&1\\0&0&0\\0&0&0\\0&0&1\\1&0&0\\0&1&0\\0&0&0\\0&0&1
\end{bmatrix}.$$

By checking the row vector of $\ti{F}_{t_1}$ (respectively, $\ti{F}_{t_2}$), we can see that the intersections of all one-dimensional error spaces with the message space are $\{\bzero\}$. This implies that the minimum distance of this code at $t_1$ (respectively, $t_2$) is 2. This shows that $\{ \f_e: e\in E \}$ constitutes a global description of a two-dimensional $\mathbb{F}_3$-valued network MDS code over the network $G$.

Further we can choose an one-dimensional $\mathbb{F}_3$-valued column vector $\vk=k=1$. Then after a simple calculation, we have $$k^{(\w-1)}_{d'_1,1}(\vk)=k^{(\w-1)}_{d'_1,4}(\vk)=2,$$ $$k^{(\w-1)}_{d'_1,2}(\vk)=k^{(\w-1)}_{d'_1,3}(\vk)=k^{(\w-1)}_{d'_1,5}(\vk)=1,$$
$$k^{(\w-1)}_{3,6}(\vk)=k_{3,6}=1, \mbox{  and  } k^{(\w-1)}_{3,7}(\vk)=k_{3,7}=1,$$
and the $(\w-1)$-dimensional decoding matrices are
$$\ti{F}^{(\w-1)}_{t_1}(\vk)=\begin{bmatrix}2&1&1\\1&0&0\\0&1&0\\0&0&1\\0&0&0\\0&0&0\\0&0&1\\0&0&0
\end{bmatrix}, \mbox{   and   }
\ti{F}^{(\w-1)}_{t_2}(\vk)=\begin{bmatrix}2&1&1\\0&0&0\\0&0&0\\0&0&1\\1&0&0\\0&1&0\\0&0&0\\0&0&1
\end{bmatrix}.$$
Further, by checking the row vectors of $\ti{F}^{(\w-1)}_{t_1}(\vk)$ (respectively, $\ti{F}^{(\w-1)}_{t_2}(\vk)$), we can see that the intersections of all two-dimensional error spaces with the message space are $\{\bzero\}$. This implies that the minimum distance of this code at $t_1$ (respectively, $t_2$) is 3. Therefore, $\{ \f^{(\w-1)}_e(\vk): e\in E \}$ constitutes an one-dimensional $\mathbb{F}_3$-valued network MDS code and the local encoding kernels at all internal nodes are the same as that of $\{ \f_e:e\in E \}$ over the network $G$.
\end{eg}

\section{Performance Analysis}

In this section, we will focus on the performance of our proposed algorithm for constructing variable-rate network MDS codes in different aspects including the field size, the time complexity of the algorithm, the encoding complexity at the source node, and the decoding methods.

First, recall that Yang \textit{et al.} \cite{Yang-refined-Singleton} proposed two algorithms for constructing network MDS codes and both of them design the codebook at the source node and local encoding kernels separately. The first one needs to find a codebook based on a given set of local encoding kernels, and the second one needs to find a set of local encoding kernels based on a given classical error-correcting code at the source node satisfying a certain minimum distance requirement as the codebook. Hence, it seems likely that these two algorithms might solve this variable-rate network error correction problem. However, by a detailed analysis below, they are either non-feasible or inefficient for solving the problem. To be specific, for the second one, the design of the set of local encoding kernels is based on a given classical error-correcting code, say $\mathcal{C}$, at the source node, so the local encoding kernels are different for the distinct classical error-correcting codes with distinct information rates at the source node. Mathematically, for each updating channel $e$, where $e$ is the edge appended to the graph at the $i$th iteration, $i>0$, let $\bk_e=\begin{bmatrix} k_{d,e}: & d\in E^{i-1}\end{bmatrix}^\top$ be an $(|Out(s)|+i-1)$-dimensional column vector consisting of all local encoding coefficients $k_{d,e}$ for the channels $d,e$, where $k_{d,e}=0$ if $d$ and $e$ are not adjacent, and $E^{i-1}$ is the set of channels in the $(i-1)$-th subnetwork $G^{i-1}$ of $G$.  Note that $\bk_e$ has to be chosen to satisfy the following \textit{feasible condition}, that is,
$$\left(F_t^i(X,-Z^i)\right)^{\backslash\mL}\neq \bzero$$
for all combinations of
\begin{description}
  \item[C1)] $t\in T$;
  \item[C2)] $\mL\subset \{1,2,\cdots,r_t\}$ with $0\leq |\mL| \leq d_t-1$;
  \item[C3)] nonzero $ X\in \mathcal{C}\subseteq \mF^{|Out(s)|}$;
  \item[C4)] error vector $Z^i$ with $w_H(Z^i)\leq d_t-1-|\mL|$;
\end{description}
where $F_t^i(X,-Z^i)$ represents the output of the channels in $In(t)$ for input $X\in \mathcal{C}$ and error vector $-Z^i$, in the $i$th subnetwork $G^i$  corresponding to the $i$th iteration, the designed rank of the matrix $F_{s,t}$ to be introduced in (\ref{F_st}) below is $r_t$ and the designed minimum distance is $d_t$ for each sink $t\in T$ (refer to \cite{Yang-refined-Singleton}\cite{Yang-thesis} for more details).
So it is easily seen that $\bk_e$ depends on some initial parameters including the given algebraic code $\mathcal{C}$, the minimum distance $d_t$, the rank $r_t$ of the matrix $F_{s,t}$, and so on, which further depend on the information rate $\w$. This implies that it is impossible to use Yang \textit{et al.}s' Algorithm 2 to construct variable-rate network MDS codes.

The first algorithm needs to find a codebook at the source node after a set of local encoding kernels is given. Thus, it seems likely that it is feasible to design variable-rate network MDS codes to solve this variable-rate problem. However, for the first algorithm, as described by Yang \textit{et al.} \cite{Yang-refined-Singleton}, they just give a method to find the proper codebook at the source node and the part of constructing local encoding kernels makes use of the existing Jaggi \textit{et al.}s' algorithm \cite{co-construction} directly. Actually, the way of Jaggi \textit{et al.}s' algorithm to obtain linear network codes is to construct global encoding kernels for all channels one by one from the source node $s$ to each sink $t\in T$, including all outgoing channels of the source node. In other words, it designs the matrix $M\triangleq \begin{bmatrix}f_e:& e\in E\end{bmatrix}$, where $f_e$ is the global encoding kernel of channel $e$. Further, each sink node $t\in T$ can use the corresponding decoding matrix:
\begin{align*}
M_t\triangleq MA_{In(t)}^\top=\begin{bmatrix} f_e: & e\in In(t) \end{bmatrix},
\end{align*}
where we use $A_{\p}$ to denote a $|\p|\times |E|$ matrix with $\p$ being a collection of channels, to be specific, $A_{\p}=[A_{d,e}]_{d\in \p,e\in E}$ satisfying
\begin{align*}
A_{d,e}=\begin{cases}1, & d=e,\\
                     0, & \text{otherwise.}\end{cases}
\end{align*}
Particularly, $A_{In(t)}=[A_{d,e}]_{d\in In(t),e\in E}$ and $A_{Out(s)}=[A_{d,e}]_{d\in Out(s),e\in E}$. Thus, by \cite{Koetter-Medard-algebraic} (also see \cite{Zhang-book}\cite{Yeung-book}), it is not difficult to obtain that
\begin{align}\label{1}
M_t=M\cdot A_{In(t)}^\top=K_s\cdot A_{Out(s)}\cdot(I-K)^{-1}\cdot A_{In(t)}^\top= K_s\cdot F_{s,t},
\end{align}
where
\begin{align}\label{F_st}
F_{s,t}\triangleq A_{Out(s)}\cdot(I-K)^{-1}\cdot A_{In(t)}^\top;
\end{align}
$K=[k_{d,e}]_{d\in E,e\in E}$ is the system transfer matrix (also called one-step transformation matrix) of size $|E|\times |E|$ with $k_{d,e}$ being the local encoding coefficient for the adjacent pair $(d,e)$ of channels, and $k_{d,e}=0$ otherwise; $K_s=[k_{d,e}]_{d\in In(s),e\in Out(s)}$ is the local encoding kernel at the source node, and $I$ represents an $|E|\times|E|$ identity matrix.
Recall that, Yang \textit{et al.}s' Algorithm 1 first needs to construct a set of local encoding kernels satisfying $\Rank(F_{s,t})=r_t$ for each sink node $t\in T$. But, together with the equality (\ref{1}), it seems that it is not feasible to apply Jaggi \textit{et al.}s' algorithm directly. To be specific, by Jaggi \textit{et al.}s' algorithm, one obtains decoding matrices $M_t$ for all sink nodes $t\in T$. But only from $M_t$, it is difficult to find a matrix $K_s$ such that $M_t=K_s\cdot F_{s,t}$ and $F_{s,t}$ satisfies $\Rank(F_{s,t})=r_t$ for each sink node $t\in T$. So in order to apply Yang \textit{et al.}s' Algorithm 1 to solve the variable-rate problem it is necessary to design a new algorithm or modify Jaggi \textit{et al.}s' algorithm to achieve the above requirements, that is, construct local encoding kernels at all internal nodes such that $\Rank(F_{s,t})=r_t$ for each $t\in T$. We believe that modifying Jaggi \textit{et al.}s' algorithm supposedly makes sense. Furthermore, even assuming that all local encoding kernels at internal nodes satisfying the condition that $\Rank(F_{s,t})=r_t$ for each sink node $t\in T$ are given, our proposed algorithm still has many advantages in different aspects such as the size of base finite field, the time complexity of the algorithms, the encoding complexity at the source node, and the decoding algorithms. In the following, we show the detailed discussion in order to characterize the performance analysis of our algorithms.

\subsection{Field Size}

From \cite{Guang-MDS}, we have known that the required field size of our algorithm for constructing a network MDS code is smaller (in some cases much smaller) than that of Yang \textit{et al.}s' algorithms. If the variable-rate network MDS coding is considered simultaneously, the required field size of our algorithm is still smaller (also in some cases much smaller) than that of Yang \textit{et al.}s' algorithms.

Without loss of generality, we consider two variable-rate network MDS codes with respective information rates $\w$ and $\w-1$. As stated in the last section, we have obtained that if the base field size:
$$|\mF|>\max\Big\{ \sum_{t\in T}|R_t(\dt)|,\ \sum_{t\in T}|R_t(\dt+1)| \Big\},$$
then, applying our algorithm, we can construct two variable-rate network MDS codes with respective information rates $\w$ and $\w-1$, that is, the constructed $(\w-1)$-dimensional and $\w$-dimensional network MDS codes $\mC_{\w-1}$ and $\mC_{\w}$ have the same local encoding kernels at all internal nodes. Particularly, if $\dt+1\leq \lceil C_t/2 \rceil$, then we have $|R_t(\dt)|\leq |R_t(\dt+1)|$ from \cite[Lemma 9]{Guang-MDS}, which means that the field size satisfying $|\mF|>\sum_{t\in T}|R_t(\dt+1)|$ is enough. In fact, notice that the field size satisfying
$$|\mF|>\max\Big\{ \sum_{t\in T}|R_t(\dt)|,\ \sum_{t\in T}|Q(t)| \Big\}$$
is enough for constructing such two network MDS codes. Together with $|Q(t)|\leq |R_t(\dt+1)|$ and $|R_t(\dt)|\leq |R_t(\dt+1)|$, it follows
$$\max\Big\{ \sum_{t\in T}|R_t(\dt)|,\ \sum_{t\in T}|Q(t)| \Big\}\leq \sum_{t\in T}|R_t(\dt+1)|.$$
This implies that the left hand side of the above inequality is big enough for the required field size for the existence of $(\w-1)$-dimensional network MDS codes, which usually is smaller than the previous result $\sum_{t\in T}|R_t(\dt+1)|$ proposed in \cite{Guang-MDS}.

On the other hand, from Theorem 10 in \cite{Yang-refined-Singleton}, in order to construct $\w$-dimensional network MDS code, the required filed size is not less than $\sum_{t\in T}{|E| \choose \dt}$. Further, for constructing an $(\w-1)$-dimensional network MDS code with the same local encoding kernel at the internal nodes, the required field size is not less than
$$\sum_{t\in T}{|E| \choose C_t-(\w-1)}=\sum_{t\in T}{|E| \choose \dt+1}.$$
Combining the above, the required base field size of Yang \textit{et al.}s' Algorithm 1 satisfies:
$$|\mF|>\max\left\{ \sum_{t\in T}{|E| \choose \dt},\ \sum_{t\in T}{|E| \choose \dt+1} \right\}.$$
In particular, if $\dt+1\leq \lfloor |E|/2 \rfloor$, then we deduce $|\mF|>\sum_{t\in T}{|E| \choose \dt+1}$. In addition, Lemma 6 in \cite{Guang-MDS} shows that $$\sum_{t\in T}|R_t(\dt)|\leq \sum_{t\in T}{|E| \choose \dt}$$ and $$\sum_{t\in T}|R_t(\dt+1)|\leq \sum_{t\in T}{|E| \choose \dt+1},$$ which indicates that our algorithm needs smaller field size than Yang \textit{et al.}s'.

\begin{eg}
Let $G$ be a combination network \cite[p.26]{Zhang-book}\cite[p.450]{Yeung-book} with parameters $N=6$ and $k=4$. To be specific, $G$ is a single source multicast network with $N=6$ internal nodes, where there is one and only one channel from the source node $s$ to each internal node, and arbitrary $k=4$ internal nodes are connective with one and only one sink node, which implies that there are totally ${6\choose 4}=15$ sink nodes. Thus, for $G$, we know that $|J|=6$, $|T|={6 \choose 4}=15$, and $|E|=6+4\times{6\choose 4}=66$. It is evident that the minimum cut capacity $C_t$ between $s$ and any sink node $t$ is $4$. For example, Fig. \ref{fig_cn} shows a combination network with $N=3,k=2$.
\begin{figure}[!htb]
\centering
\begin{tikzpicture}
[->,>=stealth',shorten >=1pt,auto,node distance=2cm,
                    thick]
  \tikzstyle{every state}=[fill=none,draw=black,text=black,minimum size=7mm]
  \tikzstyle{place}=[fill=none,draw=white,minimum size=0.1mm]
  \node[state]         (s)                 {$s$};
  \node[state]         (i_2)[below of=s]   {$i_2$};
  \node[state]         (i_1)[left of=i_2]  {$i_1$};
  \node[state]         (i_3)[right of=i_2] {$i_3$};
  \node[state]         (t_1)[below of=i_1] {$t_1$};
  \node[state]         (t_2)[below of=i_2] {$t_2$};
  \node[state]         (t_3)[below of=i_3] {$t_3$};
\path
(s) edge           node {} (i_1)
    edge           node {} (i_2)
    edge           node {} (i_3)
(i_1) edge node{}(t_1)
      edge node{}(t_2)
(i_2) edge node{}(t_1)
      edge node{}(t_3)
(i_3) edge node{}(t_2)
      edge node{}(t_3);
\end{tikzpicture}
\caption{Combination Network with $N=3,k=2$.}
\label{fig_cn}
\end{figure}
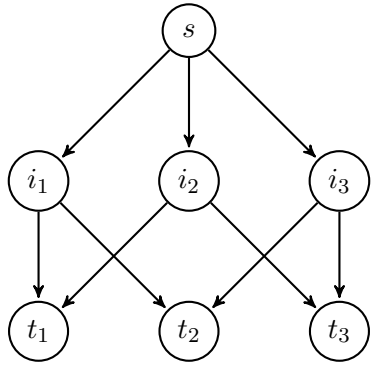
Furthermore, let the information rates be $\w=2$ and $\w=1$, and thus $\dt^{(\w=2)}=C_t-2=2$ and $\dt^{(\w=1)}=C_t-1=3$, respectively. Therefore, for each sink node $t\in T$, one has
\begin{align*}
|R_t(\dt^{(\w=2)})|&=|R_t(2)|=2^2\cdot{4 \choose 2}=24,\\
|R_t(\dt^{(\w=1)})|&=|R_t(1)|=2^3\cdot{4 \choose 3}=32,
\end{align*}
which further leads to
\begin{align*}
\sum_{t\in T}|R_t(\dt^{(\w=2)})|&=15\times 24=360,\\
\sum_{t\in T}|R_t(\dt^{(\w=1)})|&=15\times 32=480.
\end{align*}
So the field size satisfying $|\mF|>480$ is enough for our proposed algorithm.

On the other hand, we further have
$$\sum_{t\in T}{|E| \choose \dt^{(\w=2)}}=\sum_{t\in T}{66 \choose 2}=32175,$$
and
$$\sum_{t\in T}{|E| \choose \dt^{(\w=1)}}=\sum_{t\in T}{66 \choose 3}=45760.$$
Thus, Yang \textit{et al.}s' algorithm needs the field size satisfying $|\mF|>45760$.
\end{eg}

\subsection{Time Complexity of Algorithms}

Below we will discuss the time complexity of constructive algorithms. Similarly, we still consider one representative case constructing two variable-rate network MDS codes with respective dimensions $\w$ and $\w-1$.
As discussed in \cite{Guang-MDS}, the time complexity of our used algorithm for constructing an $\w$-dimensional network MDS code is smaller than that of Yang \textit{et al.}s' two algorithms by using either the random analysis method or the deterministic analysis method. In the following, we further discuss the time complexity of constructing a variable-rate $(\w-1)$-dimensional network MDS code.

Note that the key for constructing such an $(\w-1)$-dimensional network MDS code from a given $\w$-dimensional network MDS code is to choose a proper $(\w-1)$-dimensional $\mF$-valued vector $\vk$, which has to satisfy the condition (\ref{choose-vk}). Therefore, from \cite[Lemma 8]{co-construction} and \cite[Lemma 11]{Yang-refined-Singleton}, the time complexity of our algorithm to construct such an $(\w-1)$-dimensional network MDS code from a given $\w$-dimensional network MDS code is at most
$$\mO\left( (\w-1)^3\sum_{t\in T}|Q(t)|+(\w-1)\left[ \sum_{t\in T}|Q(t)| \right]^2 \right).$$
Note that it is also the encoding time complexity at the source node by using our algorithm.

In the following, we consider Yang \textit{et al.}s' Algorithm 1. First, assume that all local encoding kernels at internal nodes are given and satisfy $\Rank(F_{s,t})=C_t$, where again recall that $C_t$ is the minimum cut capacity between the source node $s$ and the sink node $t$, and $F_{s,t}=A_{Out(s)}\cdot(I-K)^{-1}\cdot A_{In(t)}^\top$. By \cite[Theorem 10 and Algorithm 1]{Yang-refined-Singleton}, if we want to construct a network MDS code with the information rate $\w$, we have to derive $\w$ $n_s$-dimensional vectors $\bg_1,\bg_2,\cdots,\bg_\w$ in turn satisfying:
\begin{align*}
\bg_1&\notin \Dt(\bzero, C_t-\w),\\
\bg_i&\notin \Dt(\bzero, C_t-\w)+\langle \bg_1,\bg_2,\cdots,\bg_{i-1} \rangle,
\end{align*}
for each $i$, $2\leq i \leq \w$, where $n_s$ is the number of outgoing channels of the source node $s$, i.e., $n_s=|Out(s)|$, and
$$\Dt(\bzero, C_t-\w)=\{ \bg:\ \bg\in \mF^{n_s} \mbox{ satisfying } \min\{ w_H(\bZ):\ \bZ \in \mF^{|E|} \mbox{ such that } \bg F_{s,t}=\bZ G_t \}\leq C_t-\w  \}$$
with $w_H(\cdot)$ representing the Hamming weight. Further, when an $(\w-1)$-dimensional network MDS code with the same local encoding kernels is constructed, we similarly derive $(\w-1)$ $n_s$-dimensional vectors $\bg_1',\bg_2',\cdots,\bg_{(\w-1)}'$ in turn according to the same way. However, it is necessary to notice that
$$\Dt(\bzero, C_t-\w)\subseteq \Dt(\bzero, C_t-\w+1),$$
which implies that $\bg_1,\bg_2,\cdots,\bg_\w$ for $\w$-dimensional network MDS code may be useless for deriving $\bg_1',\bg_2',\cdots,\bg_{(\w-1)}'$. In other words, it has to repeat the same procedure to choose the proper vectors $\bg_1',\bg_2',\cdots,\bg_{(\w-1)}'$. This evidently increases the complexity.

From \cite{Yang-refined-Singleton}, based on the given local encoding kernels at all internal nodes, the time complexity of Yang \textit{et al.}s' Algorithm 1 for constructing such an $(\w-1)$-dimensional network MDS code is
$$\mO\left( (\w-1)n_s^3\sum_{t\in T}{|E|\choose \dt+1}+(\w-1)n_s\left[ \sum_{t\in T}{|E|\choose \dt+1} \right]^2 \right),$$
which is also the encoding time complexity at the source node by using Yang \textit{et al.}s' algorithm.

Since $|Q(t)|\leq |R_t(\dt+1)|\leq {|E| \choose \dt+1}$ for each sink node $t\in T$, it is easily seen that
\begin{align*}
&(\w-1)^3\sum_{t\in T}\big|Q(t)\big|+(\w-1)\left[ \sum_{t\in T}\big|Q(t)\big| \right]^2\\
<& (\w-1)n_s^3\sum_{t\in T}{|E|\choose \dt+1}+(\w-1)n_s\left[ \sum_{t\in T}{|E|\choose \dt+1} \right]^2,
\end{align*}
and, in general cases, the former is much smaller than the later.

In view of the above discussion, the total time complexity of our algorithm for constructing variable-rate network MDS codes is also smaller than that of Yang \textit{et al.}s' algorithm. Particularly, for our algorithm, the encoding time complexity at the source node is smaller (in general much smaller) than that of Yang \textit{et al.}s' algorithm. This time complexity is important, in particular, when the local encoding kernels at all internal nodes are fixed.

In addition, during the analysis of time complexity of these algorithms, it is assumed that any arithmetic in the base finite field is $\mO(1)$ regardless of the finite field. Actually, it is well-known that the cost of arithmetic in small field is smaller than that in a bigger one, together with the above conclusion that the size of the base field used in our algorithm is smaller than that of others, which implies that the time complexity of the proposed algorithm can be reduced further.

\subsection{Decoding Algorithms}

In \cite{Yang-refined-Singleton}, Yang \textit{et al.} just gave two decoding principles by using the concept of the minimum weight (refer to Definitions 2 and 3 in \cite{Yang-refined-Singleton}), which are similar to the minimum distance decoding principle. This minimum distance decoding problem (usually called nearest codeword problem) is known to be NP-hard for classical linear codes which can be regarded as special linear network codes. Moreover, as mentioned in \cite{zhang-correction} and \cite{Guang-MDS}, our algorithm can make use of the better and faster decoding algorithms proposed by Zhang, Yan,
and Balli in a series of papers \cite{zhang-correction}, \cite{zhang-beyond}, and \cite{Zhang-survey-paper-NEC} such as the
brute force decoding algorithm and, particularly, the statistical decoding
algorithm. For the case of decoding in packet networks \cite{zhang-beyond} and \cite{Zhang-survey-paper-NEC}, where all messages such as $X_i$ $(1\leq i \leq \w)$, $Z_e$, and $\tilde{U}_e$ $(e\in E)$ are column vectors over the base field $\mF$, all message scalar components in a packet share the same extended global encoding kernel, and the decoding principle is applied to each message scalar component of the packets, our algorithm has more advantages on decoding network error correction codes beyond the error correction capability, even beyond the minimum distance.


\section{Random Variable-Rate Network MDS Codes}

At present, as described in \cite{Yang-thesis}\cite{Yang-refined-Singleton}, there are roughly two classes of network error correction coding. One class is called \textit{coherent network error correction} if the sink nodes know the network topology as well as the network codes used in transmission. Otherwise, the network error correction without this assumption is called \textit{noncoherent network error correction}. Actually, coherent and noncoherent transmissions for network coding are analogous to the coherent and noncoherent transmissions for multiple antenna channels in wireless communication. When using the deterministic construction of linear network codes such as \cite{Li-Yeung-Cai-2003} and \cite{co-construction}, the network transmission is usually regarded as coherent, and when using random network coding such as \cite{Ho-etc-random} and \cite{zhang-random}, the network transmission is usually considered to be noncoherent. Here the main idea of random network coding is that when a node (maybe the source node $s$) receives the messages from its all incoming channels, for each outgoing channel, it randomly and uniformly picks the encoding coefficients from the base field $\mF$, uses them to encode the received messages, and transmits the encoded messages over the outgoing channel. In other words, the local coding coefficients $k_{d,e}$ are independently and uniformly distributed random variables taking values in the base field $\mF$.
However, it is possible to use noncoherent transmission for deterministicly constructed linear network codes and use coherent transmission for randomly constructed linear network codes.

When the noncoherent network error correction is under consideration, for the problem discussed in this paper, the deterministic constructive algorithm may not be used since the network topology is unknown. So the above random method is also applied to noncoherent network error correction, and the linear network error correction codes constructed by this method are called random linear network error correction codes. Further, we obtain the following result.

\begin{thm}
Consider noncoherent network error correction coding on a single source multicast network $G$. Using random method to construct two variable-rate network MDS codes with respective dimensions $\w$ and $\w-1$, then the success probability $Pr(\mC_{\w}\cap\mC_{\w-1})$ for constructing such two codes satisfies:
\begin{align*}
Pr(\mC_{\w}\cap\mC_{\w-1})\geq \left[1-\frac{\sum_{t\in T}|Q(t)|}{|\mF|}\right]\cdot\left[1-\frac{\sum_{t\in T}|R_t(\dt)|}{|\mF|-1}\right]^{|J|+1},
\end{align*}
where again $\dt=C_t-\w$, and $J$ is the set of internal nodes in $G$. This further indicates two variable-rate network MDS codes with respective dimensions $\w$ and $\w-1$ can be constructed with high probability close to one by random method, if the size of the base field $\mF$ is sufficiently large.
\end{thm}
\begin{IEEEproof}
By \cite[Theorem 11]{Guang-MDS}, we know the probability $Pr(\mC_\w)$ that $\w$-dimensional network MDS codes are constructed by the random method is lower bounded by:
\begin{align}\label{pr_C_w}
Pr(\mC_{\w})\geq \left[1-\frac{\sum_{t\in T}|R_t(\dt)|}{|\mF|-1}\right]^{|J|+1}.
\end{align}
Together with $Pr(\mC_{\w}\cap\mC_{\w-1})=Pr(\mC_{\w})Pr(\mC_{\w-1}|\mC_{\w})$, it suffices to take the probability $Pr(\mC_{\w-1}|\mC_{\w})$ into account.

We randomly and uniformly pick an $(\w-1)$-dimensional column vector $\vec{k}$ from $\mF^{\w-1}$, i.e., $\vk$ is a uniformly distributed random vector taking values in $\mF^{\w-1}$. By Lemma \ref{lem_5}, it follows that, if
$$\vk\in \mF^{\w-1}\backslash \cup_{t\in T}\cup_{\p\in Q(t)}K(t,\p),$$
then $\{\f_{e}^{\w-1}(\vk):\ e\in E\}$ constitutes an $(\w-1)$-dimensional network MDS code and its local encoding kernel at each non-source node are the same as that of $\mathbf{C}_\w$. Thus, we will focus on the probability
\begin{align}\label{pr_C_w-1}
Pr(\mC_{\w-1}|\mC_{\w})\geq P(\vk)\triangleq Pr(\vk\in \mF^{\w-1}\backslash \cup_{t\in T}\cup_{\p\in Q(t)}K(t,\p)).
\end{align}
It is not difficult to obtain
\begin{align}
P(\vk)=&Pr(\vk\in \mF^{\w-1}\backslash \cup_{t\in T}\cup_{\p\in Q(t)}K(t,\p))\nonumber\\
=&\frac{|\mF^{\w-1}\backslash \cup_{t\in T}\cup_{\p\in Q(t)}K(t,\p)|}{|\mF|^{\w-1}}\nonumber\\
\geq&\frac{|\mF|^{\w-1}-\sum_{t\in T}\sum_{\p\in Q(t)}|\mF^{\w-1}\cap K(t,\p)|}{|\mF|^{\w-1}}\nonumber\\
=&1-\frac{\sum_{t\in T}|Q(t)|}{|\mF|}.\label{pr_k}
\end{align}
Combining the inequalities (\ref{pr_C_w}), (\ref{pr_C_w-1}) and (\ref{pr_k}), one obtains a lower bound on the success probability:
$$Pr(\mC_{\w}\cap\mC_{\w-1})\geq \left[1-\frac{\sum_{t\in T}|R_t(\dt)|}{|\mF|-1}\right]^{|J|+1}\cdot\left[1-\frac{\sum_{t\in T}|Q(t)|}{|\mF|}\right].$$

Furthermore, it is not difficult to see that $Pr(\mC_{\w})\rightarrow 1$ and $P(\vk)\rightarrow 1$ as $|\mF|\rightarrow \infty$ from (\ref{pr_C_w}) and (\ref{pr_k}), respectively. Therefore, for sufficiently large base field $\mF$, an $\w$-dimensional and an $(\w-1)$-dimensional network MDS codes with the same local encoding kernel at each non-source node can be constructed by random method with high probability close to one. This accomplishes the proof.
\end{IEEEproof}

Together with Corollary \ref{cor-8}, the above theorem leads to the following corollary immediately.
\begin{cor}
Using random method to construct two variable-rate network MDS codes with respective dimensions $\w$ and $\w-1$, then the success probability $Pr(\mC_{\w}\cap\mC_{\w-1})$ for constructing such two codes satisfies:
\begin{align*}
Pr(\mC_{\w}\cap\mC_{\w-1})\geq \left[1-\frac{\sum_{t\in T}{ |E|\choose C_t-\w+1}}{|\mF|}\right]\cdot\left[1-\frac{\sum_{t\in T}{|E| \choose C_t-\w}}{|\mF|-1}\right]^{|J|+1},
\end{align*}
and further for general cases with $C_t\leq \lfloor \frac{|E|}{2}\rfloor$ for all sink nodes $t\in T$,
\begin{align*}
Pr(\mC_{\w}\cap\mC_{\w-1})\geq \left[1-\frac{\sum_{t\in T}{|E| \choose C_t-\w+1}}{|\mF|-1}\right]^{|J|+2},
\end{align*}
where again $J$ is the set of internal nodes in $G$.
\end{cor}

For constructing a family of variable-rate network MDS codes by the random method, we similarly have the following corollary.
\begin{cor}
A family of variable-rate network MDS codes can be constructed with high probability close to one by the random method, if the size of the base field $\mF$ is sufficiently large.
\end{cor}

In \cite{Fong-Yeung-variable-rate}, the authors proposed a further research problem that is the performance analysis of randomly designed codes for variable-rate linear network coding. Actually, the discussions in this section analyze the performance of randomly designed network MDS codes for our variable-rate network error correction problem, which is more complicated than variable-rate linear network coding problem. Therefore, our analysis method also can be applied to characterize the performance of randomly designed variable-rate linear network codes.

\section{Conclusion}
In network communication, the source often transmits the messages at several different information rates within a session. When both information transmission and network error correction are under consideration, linear network error correction MDS codes are expected to be used for these different rates. In this paper, we propose a more efficient scheme for this purpose than using the known algorithms to construct network MDS code for each rate. In addition, these network MDS codes designed by the proposed scheme have the same local encoding kernels at all internal nodes. This saves the storage space for each internal node and resources and time for the transmission.

Some interesting problems in this direction remain open. For instance, we can also consider a family of variable-rate general linear network error correction codes with certain error correction capacity instead of network MDS codes, partly because the field size required by general linear network error correction codes is smaller than that of network MDS codes.


\end{document}